\documentclass[letterpaper, 10 pt, conference]{ieeeconf}  

\IEEEoverridecommandlockouts                              
\usepackage{times}
\usepackage{graphicx} 

\usepackage{amsmath,amssymb,amsfonts} 
\usepackage{amsthm}
\usepackage{float}
\newtheorem{definition}{Definition}
\newtheorem{cor}{Corollary}
\newtheorem{expm}{Example}
\newtheorem{remark}{Remark}
\newtheorem{lemma}{Lemma}
\newtheorem{thm}{Theorem}
\usepackage{fourier}
\usepackage{subcaption}
\usepackage{xcolor}
\usepackage{hyperref}
\usepackage{verbatim}
\usepackage{caption}
\usepackage{cite}
\usepackage{subcaption}
\usepackage[skip=3pt]{caption} 

\renewcommand{\baselinestretch}{0.95}
\title{\LARGE \bf
Leveraging Network Topology in a Two-way Competition for Influence in the Friedkin-Johnsen Model}

\author{Aashi Shrinate$^{1}$, \IEEEmembership{Student member, IEEE}, Aravind Seshadri$^2$, Twinkle Tripathy$^3$, \IEEEmembership{Senior Member, IEEE} 
\thanks{$^{1}$Aashi Shrinate is a research scholar, $^2$ Aravind Seshadri is an undergraduate student and $^{3}$ Twinkle Tripathy is an Assistant Professor in the Control and Automation specialization of the Department of Electrical Engineering, Indian Institute of Technology Kanpur, Kanpur, Uttar Pradesh, India, 208016. Email: {\tt\small aashis21@iitk.ac.in , aravinds21@iitk.ac.in and ttripathy@iitk.ac.in}.}
}

\begin{document}

\maketitle
\thispagestyle{empty}
\pagestyle{empty}

\begin{abstract}
In this paper, we consider two stubborn agents who compete 
for `influence' over a strongly connected group of agents. 
This framework represents real-world contests, such as competition among firms, two-party elections, and sports rivalries, among others. Considering stubbornness of agents to be an immutable property, we utilise the network topology alone to increase the influence of a preferred stubborn agent.  
We demonstrate this on a special class of strongly connected networks by identifying the supporters of each of the stubborn agents in such networks. Thereafter, we present sufficient conditions under which a network perturbation always increases the influence of the preferred stubborn agent. A key
advantage of the proposed topology-based conditions is that they hold independent of the edge weights in the network. Most importantly, we assert that there exists a sequence of perturbations that can 
make \textit{the lesser influential stubborn agent more influential}.
Finally, we demonstrate our results over the Sampson's Monastery dataset.
\end{abstract}

\section{INTRODUCTION}

An individual generally forms opinions and takes decisions on crucial financial, electoral, and social issues based on interactions with their peers. In these interactions, often some peers are influential and have a substantial impact on the final decision. In the literature on social networks, `centrality measures' are often used to identify such agents and quantify their impact on others.
The author in \cite{friedkin1991theoretical} asserts that a centrality measure is useful in quantifying influence in a social network only if it arises from the underlying social process.  
While the opinion dynamics models proposed in the literature \cite{10.2307/2118364,degroot1974reaching,FJ_Model,rainer2002opinion} capture various aspects on social interactions, 
the Friedkin-Johnsen (FJ) model \cite{FJ_Model} is widely popular due to its analytical tractability and proven empirical validity.
In the FJ framework, the agents are heterogeneous with varying degrees of stubbornness towards their internal biases. For the FJ model,
the notion of \textit{influence centrality}, proposed in \cite{Community_Cleavage}, quantifies the impact of stubborn agents on the final opinion of the group. 

Influence centrality depends on both the stubborn behaviour and the underlying network topology. 
Often, the influence centrality of an agent is also referred to as its social power \cite{Tian2022}. Several works on social power \cite{Xu_stubbornness,Lingfei_wang,Multi_dim_concatenated} demonstrate that an increase in the stubborn behaviour of an agent results in the agent achieving a higher social power in the group.
However, the stubborn behaviour of an agent is often its inherent property, which may not change arbitrarily \cite{rebalancing_feeds}. Then,
suitable modifications to the underlying network topology can be an alternate tool to desirably shape the influence centrality. 

\textit{Related Literature:}  
Recent works \cite{gionis2013opinion,musco2018minimizing,reduce_conflict,reduce_harm,agents_compete_games} examine the impact of modifying network topology on final opinions and associated measures such as polarisation, conflict and influence centrality \textit{etc.} 
In the seminal paper \cite{gionis2013opinion}, the authors show that the influence centrality of a stubborn agent is equal to the probability of a random walk reaching the absorbing node formed by the stubborn agent in the augmented interaction network. Since then, several proposed algorithms employ targeted edge addition, removals, \textit{etc.,} to mitigate polarisation \cite{musco2018minimizing}, conflict \cite{reduce_conflict}, exposure to malicious content \cite{reduce_harm} in a social network.
 The algorithm presented in \cite{reduce_harm} selects a webpage on the internet and exchanges its out-neighbour with another webpage to reduce the probability of a random surfer encountering harmful content. Since the web surfing model is analogous to the FJ model (see \cite{proskurnikov2016pagerank} for details), this problem is equivalent to increasing the influence centrality of informative web pages over malicious ones. Here, weak network connectivity allows the existence of safe webpages (with less exposure to harmful content) that are substituted with existing neighbours of nodes to reduce the influence of harmful webpages.
 In \cite{agents_compete_games}, the authors consider two stubborn agents competing for influence centrality in
 a zero-sum game with each stubborn agent suitably modifying its interconnections with the rest of the agents to maximise its influence centrality. 

In this paper, we consider a group of $n$ agents whose opinions evolve by the FJ model, with two stubborn agents who compete for influence over the group. Like \cite{rebalancing_feeds}, we consider stubbornness to be an inherent property and employ only edge modifications of the form $(a,b,d)$ to improve influence centrality. The modification $(a,b,d)$ simulates \textit{feed alterations on social media} and consists of two steps: the addition of edge  $(a,b)$ followed by a reduction in edge weight of an existing edge $(d,b)$. 
Unlike \cite{reduce_harm} and \cite{agents_compete_games}, the agents form a strongly connected graph. 
Due to strong connectivity, we show that an edge modification that adds an edge from the stubborn agent can sometimes even reduce its influence centrality making the analysis more complex. 
Thereafter, by leveraging certain topological properties of the network, we present sufficient conditions under which an edge modification \textit{always} increases the influence of a desired stubborn agent.
In this regard, our major contributions as follows:
\begin{itemize}  
 \item We present a topological characterisation of the nodes  in the network to identify the `supporters' of the stubborn agents. Then, we determine the suitable conditions under which an increase in the influence centrality of a desired agent is guaranteed by making edge modifications using its supporters. The key advantage of the proposed approach is that it is independent of the edge weights in the network.
   \item We also present sufficient conditions under which a sequence of edge modifications transforms a less influential stubborn agent into a more influential one.
    \item In \cite{agents_compete_games}, the authors propose the increase the influence centrality of a stubborn agent 
    by adding edges emanating only from it. Our framework expands the scope of network modifications by showing the existence of a plethora of other suitable edge modifications. 


    
\end{itemize}
The organisation of the paper is as follows: Sec. \ref{sec:Prelims} introduces
 the required preliminaries. In Sec. \ref{sec:Prob_Form}, we define the edge modifications used to modify influence centrality. In Sec. \ref{sec:Key_Topo}, we present some useful graph-based properties. Sec. \ref{sec:Iden_usef} outlines the conditions for desirable changes in the influence centrality. The simulation results are presented in Sec. \ref{sec:sim_res}. In  Sec. \ref{sec:conc}, we conclude with insights into the future research directions.

\section{PRELIMINARIES}
\label{sec:Prelims}
Consider a graph $\mathcal{G}=(\mathcal{V},\mathcal{E})$ where $\mathcal{V}=\{1,2,...,n\}$ is the set of nodes and $\mathcal{E} \subseteq \mathcal{V} \times \mathcal{V}$ is the set of edges. Each node $i\in \mathcal{V}$ denotes an agent, and an edge $(i,j) \in \mathcal{E}$ represents the flow of information from agent $i$ to agent $j$. The weighted adjacency matrix of $\mathcal{G}$ is defined as $W=[w_{ij}]$ where $w_{ij}>0$ if edge $(j,i)\in \mathcal{E}$ else $w_{ij}=0$. The in-neighbours of a node $i\in V$ is defined as $N_{in}(i)=\{j:(j,i)\in \mathcal{E}\}$. 
An ordered sequence of nodes with each adjacent pair forming an edge is a path. A simple path is one in which none of the nodes are repeated. A cycle is a path where the initial and final nodes coincide. We define the path gain of a path as the product of the edge weights of all edges along the path. 
A network is strongly connected if there exists a path from each node in the network to every other node.  

\subsection{Friedkin-Johnsen Model}
The FJ model is an opinion dynamics model that considers individuals with varying degrees of openness towards their interactions with neighbours. 
The opinion of an agent governed by the FJ model evolves as follows:
\begingroup
\setlength{\abovedisplayskip}{2pt}
\setlength{\belowdisplayskip}{2pt}
\begin{align}
\label{eqn:op_model}
\mathbf{x}(k+1)=(I-\beta)W\mathbf{x}(k) + \beta \mathbf{x}(0)  
\end{align}
\endgroup

where $\mathbf{x}(k)=[x_1(k),...,x_n(k)] \in \mathbb{R}^n$ denotes the opinion of $n$ agents in $\mathcal{G}$ at the $k^{th}$ instance, $W$ is the weighted adjacency matrix, and $\beta=diag(\beta_1,...,\beta_n)$ is a diagonal matrix with $\beta_i \in [0,1]$ denoting the degree of stubbornness of agent $i \in \mathcal{V}$. An agent $i \in \mathcal{V}$ is a stubborn if $\beta_i>0$. The matrix $W$ is row-stochastic.

\begin{lemma}[\cite{FJ_Model}]
In a strongly connected network $\mathcal{G}=(\mathcal{V},\mathcal{E})$, if the opinion of agents evolves by eqn. \eqref{eqn:op_model} and $\beta_i>0$ for an agent $i \in \mathcal{V}$, then the final opinion converge as follows:
\begingroup
\setlength{\abovedisplayskip}{2pt}
\setlength{\belowdisplayskip}{2pt}
\begin{align}
\label{eqn:steady_state_op}
    \mathbf{x}^{*}=(I_n-(I_n-\beta)W)^{-1}\beta \mathbf{x}(0) 
\end{align}
\endgroup

Here, $\mathbf{x}^{*}=[x_1^*,...,x_n^*]$ denotes the final opinion of agents. 
\end{lemma}

\addtolength{\textheight}{-3cm}   

\section{PROBLEM FORMULATION}

\label{sec:Prob_Form}
In a society, often there are individuals with high credibility who wield significant power over the beliefs and actions of a broad audience. Such agents are referred to as influential agents. Formally, an agent in a network is \textit{influential} if its initial stand (opinion) contributes to the final outcome (opinion) achieved after discussions in a social network. In this paper, we study the evolution of opinions in social networks by the FJ model owing to its analytical tractability and performance on both small and large datasets. The FJ model attributes the stubborn behaviour as a characteristic of an influential agent. 
\begin{definition}[\cite{Community_Cleavage}]Under the evolution of opinions by \eqref{eqn:op_model} in a network $\mathcal{G}$, the influence centrality $c_i$ quantifies the contribution of the stubborn agent $i's$ initial opinion in the mean final opinion of the agents in $\mathcal{G}$. 
 \begingroup
\setlength{\abovedisplayskip}{2pt}
\setlength{\belowdisplayskip}{2pt}
\begin{align}
\label{eqn:influence_centrality}
    \mathbf{c}=\frac{P^T \mathbb{1}_n}{n}
\end{align}
\endgroup
\end{definition}
where $\mathbf{c}=[c_1,...,c_n]$ denotes the influence centrality measure of $n$ agents and $P=(I_n-(I_n-\beta)W)^{-1}\beta$. 
The influence centrality 
of an agent depends on both the stubbornness and the underlying network topology. By definition, $\mathbb{1}^{T}\mathbf{c}=1$. Note that this definition of influence centrality holds when each non-stubborn agent in the network has a path to a stubborn agent. 




\subsection{Problem formulation}
\label{subsec:pf}
Consider a strongly connected network $\mathcal{G}$ with two stubborn agents $s_1$ and $s_2$ who compete for influence over the agents. The stubborn agents have $\beta_i \in (0,1)$ to maintain strong connectivity. It follows from eqn. \eqref{eqn:influence_centrality} that an increase in the influence centrality of a stubborn agent occurs at the expense of the other. Such a competition can be seen in election campaigns, market duopolies, geopolitical rivalries, \textit{etc.} While in the FJ framework, it may seem that the agents with a higher degree of stubbornness have higher influence centrality, we present the following counter-examples. 

\begin{figure}[h]
    \centering
\includegraphics[width=0.65\textwidth,height=2.2cm,keepaspectratio]{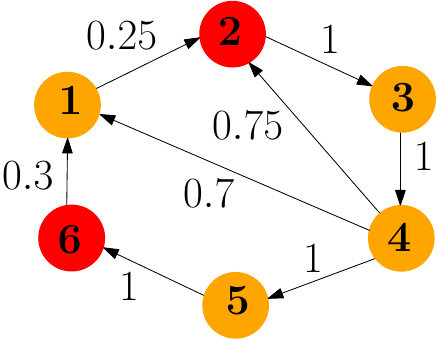}
            \caption{A network with two stubborn agents}
    \label{fig:motivating_examples}
\end{figure}

\begin{expm}
\label{expm:1}
    Consider the network shown in Fig. \ref{fig:motivating_examples} consisting of the stubborn agents $2$ and $6$. While the degree of stubbornness $\beta_6=0.8$ is higher than $\beta_2=0.2$, the influence centrality $c_6=1/3$ is lower than $c_2=2/3$.
\end{expm}

Example \ref{expm:1} shows that the network topology has a significant impact on influence centrality. \textit{Can further increasing the stubbornness of agent $6$ help?} 
\begin{expm}
\label{expm:2}
Consider the network examined in Example \ref{expm:1}. To amplify the influence of agent $6$, we increase its stubbornness to $\beta_6=0.99$ while $\beta_2$ is still $0.2$. Using eqn. \eqref{eqn:influence_centrality}, we observe that while the influence centrality of agent $6$ increases and becomes $0.4$, it is still significantly lesser than that of agent $2$ at $0.6$. 
\end{expm}
It is clear from Example \ref{expm:2} that increasing the degree of stubbornness of an agent does not always make it more influential; perturbing the network topology is sometimes the only option. In the literature, perturbations to a network are often interpreted as recommendations provided by social media networks, news, \textit{etc}., which gradually shift the opinions of the users in the network. \cite{reduce_harm}.
\textit{So, how do we perturb a network?} 


 %

\subsection{Permissible network perturbations}
\begin{definition}
\label{def:network_perturbations}
    Given distinct nodes $a$, $b$ and $d$, an edge modification $(a,b,d)$ is the addition of an edge $(a,b)$ with edge weight $w$ and the reduction of edge weight of the existing edge $(b,d)$ by $w$ ensuring that the \textit{in-degree of $b$ remains constant}
    \begingroup
\setlength{\abovedisplayskip}{2pt}
\setlength{\belowdisplayskip}{2pt}
\begin{align}
\label{eqn:weight_condition}
    w_{bd}=w+\tilde{w}_{bd}
\end{align}
\endgroup

where $w_{bd}>0$ and $\tilde{w}_{bd}>0$ are the weights of the edge $(d,b)$ before and after the edge modification, respectively. Note that $w<w_{bd}$. When an edge $(a,b)$ already exists, then $(a,b,d)$  implies an increase in edge weight of $(a,b)$ by $w$ followed by a reduction of edge weight of $(d,b)$ by $w$. 
\end{definition}

The proposed edge modification models the phenomenon of \textit{feed alteration} in social networks where the visibility of content from certain pages/influencers \textit{etc.} is increased to boost user engagement. Often, this increased visibility results in reduced occurrence of content from existing but inactive friends/pages.
In the following sections, we present the conditions under which an edge modification $(a,b,d)$ is guaranteed to increase the influence of a stubborn agent.
\section{Key Topological Properties in a Network}
\label{sec:Key_Topo}
In a network of $n$ agents with two of them being stubborn, 
the primary objective of the paper is to make one stubborn agent more influential than the other. It has been argued in \cite{rebalancing_feeds} that stubbornness is  
an intrinsic property of an agent and is not always possible to change and sometimes making suitable network perturbations is the only option as seen in Sec.  \ref{subsec:pf}.
Hence, we focus solely on the permissible network modifications introduced in Def. \ref{def:network_perturbations}. A natural question that arises is where should $a$, $b$ and $d$ lie in the network or \textit{what topological properties they must satisfy}?
\subsection{Global and local communicators}
Identifying the suitable choices for the nodes $a,b$ and $d$ is a challenging problem, specially in densely connected graphs. To make the problem tractable, we define a class of strongly connected graphs on which the analysis is conducted in this paper. 
\begin{definition}
  \label{A:gs1}  
  ($\mathbf{\mathcal{C}^1}$ \textbf{graphs, global and local communicators}): A strongly connected network $\mathcal{G}=(\mathcal{V},\mathcal{E})$ is of \textbf{type} $\mathbf{\mathcal{C}^1}$  if there is a node $m \in \mathcal{V}$ that belongs to each cycle of the network (except the self-loops). The node $m$ present in each cycle of $\mathcal{G}$ is referred to as a \textit{global communicator} while the remaining nodes in $\mathcal{G}$ are \textit{local communicators}.
\end{definition}

\begin{figure}[h]
    \centering
    \begin{subfigure}{0.23\textwidth}
        \centering
\includegraphics[width=0.95\linewidth,height=2.0cm,keepaspectratio]{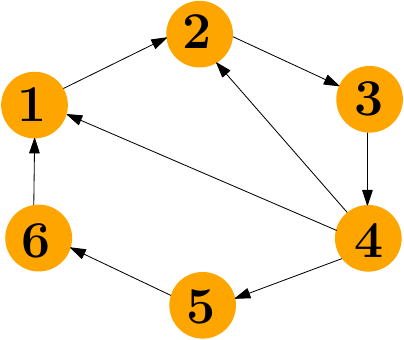}
    \caption{Network $\mathcal{G}$}
    \label{fig:gs_1_expm}
    \end{subfigure}
        \begin{subfigure}{0.23\textwidth}
        \centering
\includegraphics[width=0.95\linewidth,height=2.0cm,keepaspectratio]{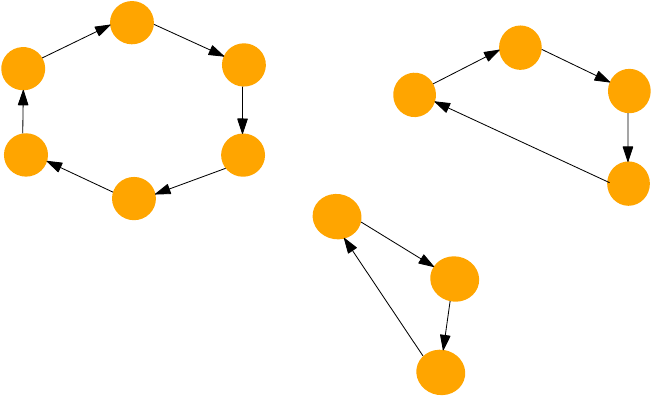}
    \caption{The simple cycles in $\mathcal{G}$.}
    \label{fig:gs_2_expm}
    \end{subfigure}
\end{figure}

In the real world, a network of type $\mathcal{C}^{1}$ can represent a political, military or business organisation that is overtly dependent on a `central leadership' participating in even the smallest decisions. A strongly connected star digraph where each node interacts only with the central node is an extreme example of a network of type $\mathcal{C}^{1}$. 
\begin{expm}
\label{expm:global_communicators}
The network $\mathcal{G}$ in Fig. \ref{fig:gs_1_expm} is strongly connected. Fig. \ref{fig:gs_2_expm} illustrates that each of the simple cycles in $\mathcal{G}$ contains nodes $2,3$ and $4$. Therefore, $\mathcal{G}$ is of type ${\mathcal{C}^1}$ and the nodes $2,3$ and $4$ are global communicators with the remaining nodes $1,5$ and $6$ being local communicators.    
\end{expm}
\vspace{-0.4cm}
\vspace{-0.4cm}

Due to the structure of type $\mathcal{C}^{1}$ graph, $m$ serves as a crucial intermediary enabling interactions among agents that are otherwise far apart in the network.
Since a global communicator $m$ in $\mathcal{G}$ passes through every cycle, we can arrange 
the nodes in $\mathcal{G}$ based on their proximity to $m$ as explained in the following lemma. 

\begin{lemma}[\cite{aashiIC}]
\label{lemma:distribution_of_nodes}
For a global communicator 
$m$, the nodes in the network  $\mathcal{G}$ are distributed into disjoint sets $\mathcal{L}_0^{m},\mathcal{L}_1^{m},...,\mathcal{L}_q^{m}$ where $q<n$. Let $m$ belong to $\mathcal{L}_0^{{m}}$. A node $j \in \mathcal{L}_z^{m}$ where $z\in\{1,2,...,q\}$ satisfies the following:
\begin{itemize}
    \item the in-neighbours of $j$ belong only from sets $\mathcal{L}_1^{m},...,\mathcal{L}_{z-1}^{m}$ and $m$,
    \item at least one in-neighbour of $j$ belongs to $\mathcal{L}_{z-1}^{m}$.
\end{itemize}
\end{lemma}
Fig. \ref{fig:distribution_nodes} illustrates the distribution of nodes of a graph of type $\mathcal{C}^{1}$ according to $m$.
Note that $m$ can be an in-neighbour or an out-neighbour of any agent in $\mathcal{G}$. It is also noteworthy that the distribution of nodes into the sets $\mathcal{L}_0^{m},\mathcal{L}_1^{m},...,\mathcal{L}_q^{m}$ is based on their proximity to $m$. Consequently, if a network $\mathcal{G}$ has two or more global communicators, multiple distributions of nodes can arise, each corresponding to a different global communicator. 

\subsection{Regions in the network $\mathcal{G}$}
In a network $\mathcal{G}$ with nodes distributed according to global communicator $m$, let stubborn agents $s_1$ and $s_2$ be such that $s_1 \in \mathcal{L}_{u}^{m}$ and $s_2 \in \mathcal{L}_{v}^{m}$ where $u\leq v$.
Based on the positioning of stubborn agents relative to $m$, we obtain the following regions in the distribution: 
\begin{itemize}
    \item  Region $1$ consists of nodes in $\mathcal{L}_0^{m},...,\mathcal{L}_{u-1}^{m}$, 
    \item Region $2$ consists of nodes in $\mathcal{L}_{u}^{m},...,\mathcal{L}_{v-1}^{m}$ and,
    \item Region $3$ consists of the nodes in  $\mathcal{L}_{v}^{m},...,\mathcal{L}_{q}^{m}$.
\end{itemize}
 \vspace{-0.35cm}
\begin{figure}[h]
    \centering    \includegraphics[width=1\linewidth,height=3.5cm,keepaspectratio]{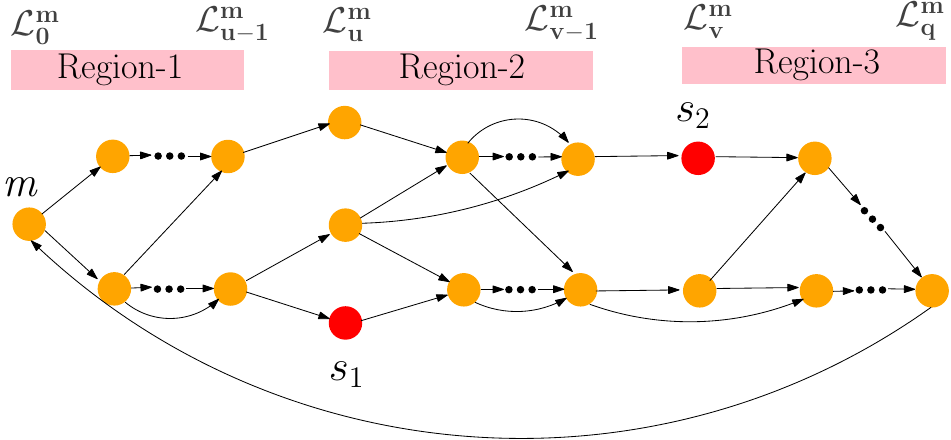}
    \caption{Regions in network $\mathcal{G}$}
    \label{fig:distribution_nodes}
\end{figure}
\vspace{-0.3cm}
To identify the nodes in $\mathcal{G}$ who are `closely connected' with each of the stubborn agents, we use the notion of direct paths defined in \cite{aashiIC}.
When the nodes in a network $\mathcal{G}$ are distributed according to the global communicator $m$, a \textit{direct path} from node $i$ to $j$ is such a path from $i $ to $ j$ which does not pass through $m$.
Based on direct paths from the stubborn agents, the nodes in $\mathcal{G}$ can be classified into the following sets: 
\begin{itemize}
   \item $\mathcal{T}_1$: a node with a direct path from only $s_1$ (not $s_2$) .
    \item $\mathcal{T}_2$: a node with a direct path from only $s_2$ (not $s_1$).
      \item $\mathcal{T}_3$: a node with a direct path from both $s_1$ and $s_2$.
    \item $\mathcal{T}_4$: a node with a direct path from neither $s_1$ nor $s_2$. 
\end{itemize}

It is evident from Fig. \ref{fig:distribution_nodes} that any node in the Region-1 is in set $\mathcal{T}_4$ as the paths originating from the stubborn agents to it must traverse through $m$. Similarly, a direct path from $s_2$ to any node in Region-2 does not exist, so it is either in $\mathcal{T}_1$ or $\mathcal{T}_4$. The nodes remaining in Region 3 can be in any aforementioned sets.

\begin{remark}
\label{rem:exceptions}
In general, we consider the stubborn agents $s_1$ and $s_2$ to be in $\mathcal{T}_1$ and $\mathcal{T}_2$, respectively and $m$ in $\mathcal{T}_3$. However, if $s_2$ has a direct path from $s_1$, then $s_2$ and each node that has a direct path from $s_2$ are in set $\mathcal{T}_3$. Consequently, $\mathcal{T}_2=\emptyset$.
     
\end{remark}

In the following section, we utilise the path-based classification of nodes to select the nodes $a,b$ and $d$ such that edge modification $(a,b,d)$ conclusively increases the influence centrality of a preferred stubborn agent.

\section{Identification of useful edge modifications}
\label{sec:Iden_usef}

The topological properties of networks of type $\mathcal{C}^{1}$, presented in Sec. \ref{sec:Key_Topo}, play a key role in the identification of suitable edge modifications. Keeping this in mind, we consider an edge modification $(a,b,d)$ to be \textit{ permissible} only if the modified network $\mathcal{G}$ remains of type $\mathcal{C}^{1}$.
The following results present the conditions under which a permissible $(a,b,d)$ increases the influence centrality of a desired stubborn agent. Prior to that, we identify the redundant modifications that do not change the influence centrality of either of the stubborn agents. Identifying such modifications helps prevent any unnecessary alterations to the network.
\begin{thm}
\label{thm:3}
Consider a network $\mathcal{G}$ of type $\mathcal{C}^1$ with two stubborn agents. The opinions of the agents in $\mathcal{G}$ evolve according to eqn. \eqref{eqn:op_model}. A permissible edge modification $(a,b,d)$ is redundant if a global communicator $m$ exists such that nodes $a$ and $d$ satisfy any one of the following conditions: 
\begin{enumerate}
    \item \textit{Equally neutral:} both $a$ and $d$ are in $\mathcal{T}_4$ or,
    \item \textit{Equally supportive:}  both $a$ and $d$ are in $\mathcal{T}_1$ (or $\mathcal{T}_2$) such that each simple path from $m$ to both $a$ and $d$ passes through $s_1$ (or $s_2$)
    \item  \textit{Equally connected:} there is a node $c$ such that each simple path from $s_1$ and $s_2$ to $a$ and $d$ traverses $c$.  
\end{enumerate}
\end{thm}
The proof of Theorem \ref{thm:3} is given in Appendix A. An edge modification $(a,b,d)$ effectively makes $a$ more interactive than $d$. Such a modification becomes redundant when there is no incentive to make this change. 
Theorem \ref{thm:3} presents the topological conditions when either both $a$ and $d$ are equally good supporters of $s_1$ or $s_2$ (Condition 2), or both are equally indifferent to them (Condition 1), making $(a,b,d)$ redundant. 
Similarly, under Condition $(3)$, both $a$ and $d$ equally well connected  $s_1$ and  $s_2$ resulting in the redundancy. Notably, 
the conditions presented in Theorem \ref{thm:3}
are independent of the distribution of the edge weights in $\mathcal{G}$, $w$ and stubborn behavior. 

\begin{thm}
\label{thm:4}
Consider a network $\mathcal{G}$ of type $\mathcal{C}^1$ with two stubborn agents. The opinions of agents evolve according to eqn. \eqref{eqn:op_model}. 
A permissible non-redundant modification $(a,b,d)$ always increases the influence centrality of $s_1$ (or $s_2$) under any one of the following topological conditions: 

\begin{enumerate}
    \item[C1.] If node $a$ satisfies the conditions: (i) $a \in \mathcal{T}_1$ (or $\mathcal{T}_2$) and (ii) each simple path from $m$ to $a$ passes through $s_1$ (or $s_2$) but node
    $d$ violates either (i) or (ii) or both.
   
      \item[C2.] If $a$ $\in \mathcal{T}4$ and $d\in \mathcal{T}_2$ (or $\mathcal{T}_1)$.
      \end{enumerate}
\end{thm}
The proof of Theorem \ref{thm:4} is given in Appendix A. Theorem \ref{thm:4} outlines the topology-based conditions under which an edge modification $(a,b,d)$ increases the influence centrality of a preferred stubborn agent.  Condition $C1$ implies that $a$ only communicates with $s_2$ while keeping $s_1$ informed. On the other hand, $d$ also receives information from $s_2$ that does not pass through $s_1$ either directly when $d \in \mathcal{T}_2$ or $\mathcal{T}_3$ or indirectly through $m$ when $d \in \mathcal{T}_1$ or  $\mathcal{T}_4$. Hence, topologically, $a$ favours $s_1$ more than $d$ does. The same effect can be achieved if $d$ favours $s_2$ more as the weight of the edge $(d,b)$ reduces. Like Theorem \ref{thm:3}, the path-based conditions presented in Theorem \ref{thm:4} are independent of the edge weights in $\mathcal{G}$ and stubbornness. 

Naturally, a question arises: \textit{given an arbitrary network of type $\mathcal{C}^1$, does a suitable modification always exist to increase the influence of either $s_1$ or $s_2$?} When we want to increase the influence centrality of $s_2$, 
we know from Remark \ref{rem:exceptions} that $\mathcal{T}_2$ is empty when a direct path from $s_1$ to $s_2$ exists. Further, certain networks of type $\mathcal{C}^{1}$ can have $\mathcal{T}_4=\emptyset$. 
Consequently, a suitable choice for $a$ such that $(a,b,d)$ increases the influence of $s_2$ does not exist. The following corollary presents an alternative for such networks.
\begin{cor}
\label{Cor:stubbornness}
Consider a network $\mathcal{G}$ of type $\mathcal{C}^{1}$ with stubborn agents $s_1$ and $s_2$ who do not have self-loops.
If $s_1$ has a direct path to $s_2$ in $\mathcal{G}$, then a permissible non-redundant edge modification $(a,b,d)$ increases the influence centrality of $s_2$ if all of the following conditions hold:
\begin{itemize}
    \item node $a$ satisfies the following: (i) $a \in \mathcal{T}_3$ and (ii) each simple path from $m$ to $a$ passes through $s_2$
    \item node
    $d$ violates either (i) or (ii) or both 
    \item $\beta_{s_2} \geq 1/2$.
\end{itemize}
\end{cor}
The proof of Corollary \ref{Cor:stubbornness} is in Appendix A. 
Corollary \ref{Cor:stubbornness} highlights 
the role of the topological position of stubborn agents because a stubborn agent $s_1\in \mathcal{L}_u^{m}$ in the lower-index set (in the distribution according to $m$) can always increase its influence centrality. However, if a direct path from $s_1$ to $s_2$ exists, the stubborn agent $s_2 \in \mathcal{L}_v^{m}$ in the higher index set can increase its influence centrality only when its degree of stubbornness is $0.5$ or higher.

\begin{thm}
\label{thm:5}
 Consider a network $\mathcal{G}$ of type $\mathcal{C}^1$ graphs with stubborn agents $s_1$ and $s_2$ without any self-loops. When the opinions of the agents in $\mathcal{G}$ evolve according to eqn. \eqref{eqn:op_model} and $\beta_{s_2} \geq 1/2$, an edge modification $(a,b,d)$ always exists which increases the influence centrality of $s_1$ or $s_2$.
\end{thm}
While the proof of Theorem \ref{thm:5} is omitted due to space constraints, it follows from a simple fact under the given conditions a node $d$ always exists such that making a stubborn agent more interactive using modification $(s_i,b,d)$ increases the $s_i$'s influence centrality for $i\in\{1,2\}$. 

An important consequence of Theorem \ref{thm:5} is that under the given conditions a sequence of suitable edge modifications always exists that can transform the \textit{least influential stubborn agent to become the most influential.} This holds because influence centrality is bounded in $(0,1)$ and we can construct a monotonically increasing sequence using Theorem \ref{thm:5}. 

\section{Simulation Results}
\label{sec:sim_res}
\begin{figure}[h]
    \centering
    \centering
\includegraphics[width=\textwidth,height=3.350cm,keepaspectratio]{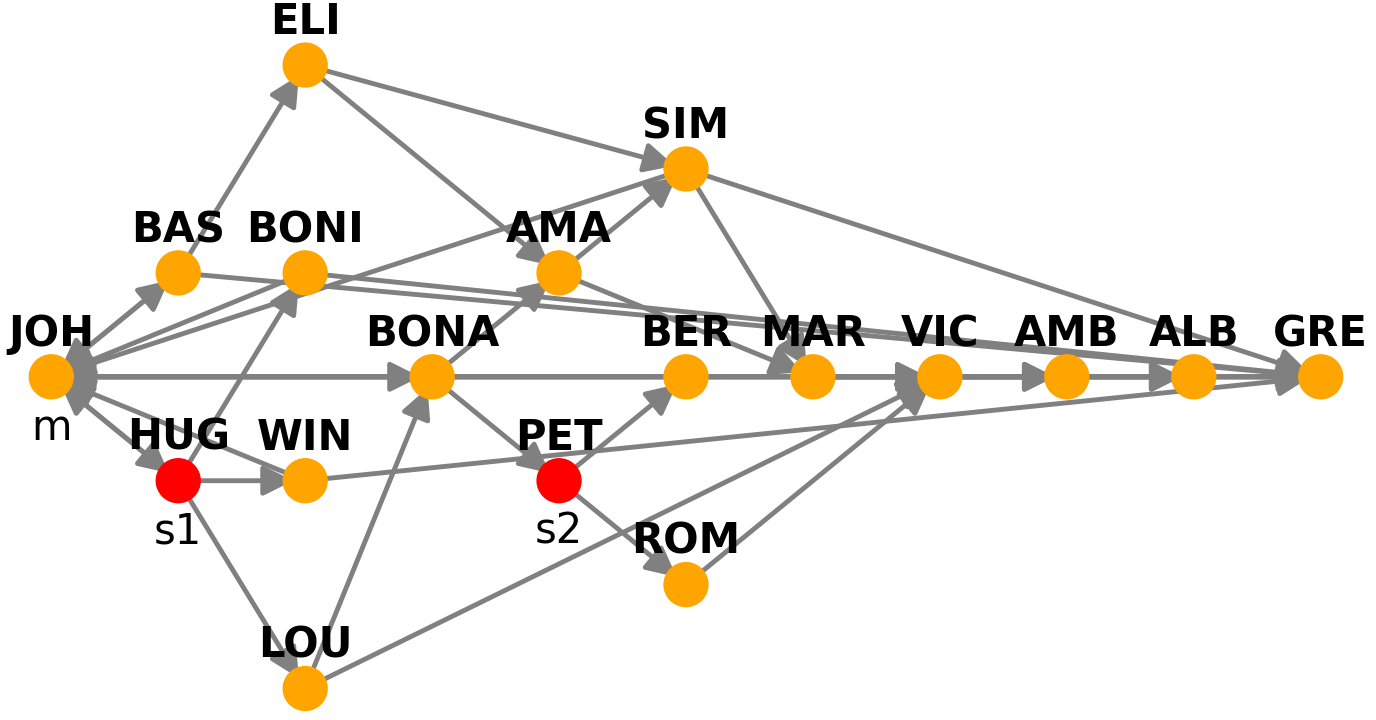} 
    \caption{The modified graph of type $\mathcal{C}^{1}$ constructed from Sampson's monastery dataset.}
    \label{fig:sampson_monastry}
\end{figure}

\begin{figure}[t]
    \centering
\includegraphics[width=\textwidth,height=3.5cm,keepaspectratio]{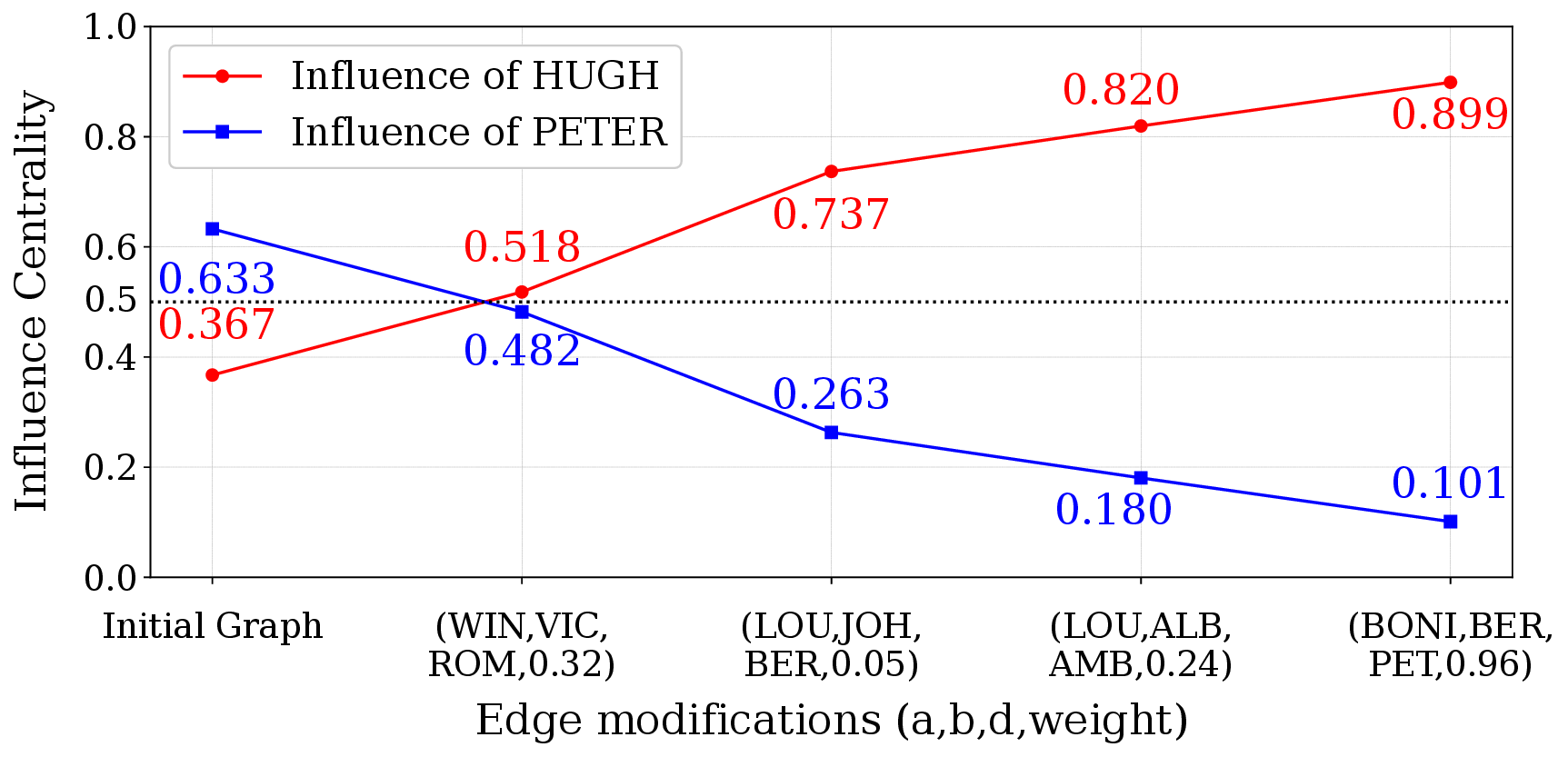} 
    \caption{Suitable edge modifications result in initially less influential Hugh becoming more influential than Peter.}
    \label{fig:influence_plot}
\end{figure}

In this section, we validate our results on the `liking' interpersonal relation network in 
Sampson monastery dataset. In the monastery, Sampson recorded two ideological factions: Young Turks and Loyal Oppositions (\cite{sampson_monks}). In this work, we consider monks Peter and Hugh, the members of Loyal Opposition and Young Turks, respectively, to be the stubborn agents who compete for ideological dominance over the group.
We modify the initial network to obtain a network of type $\mathcal{C}^{1}$ shown in Fig. \ref{fig:sampson_monastry} with John Bosco as the only global communicator.
We consider the degree of stubbornness of Peter and Hugh equal to $0.7$ and $0.1$, respectively. 
Initially, Peter has higher influence centrality than Hugh with $c_{Pet}=0.63$, and $c_{Hug}=0.37$. To increase the influence centrality of Hugh, we perform a series of edge modifications, as shown in Fig. \ref{fig:influence_plot}. Consequently, the influence centrality of Hugh increases to 0.9 and of Peter reduces to 0.1. The implementation is available in our Github Repository\footnote{\href{https://github.com/AravindS1506/Net\_Topology\_Influence.git}{https://github.com/AravindS1506/Net\_Topology\_Influence.git}}.

\vspace{-0.3cm}

\vspace{-0.4cm}
\section{CONCLUSIONS AND FUTURE WORKS}
\label{sec:conc}
This paper examines a competition for influence between two stubborn agents and proposes suitable edge modifications to improve the influence of the preferred stubborn agent. To identify the useful edge modifications, we explore the topological properties of networks of type $\mathcal{C}^{1}$ whose nodes can be distributed into disjoint sets based on their proximity with a \textit{global communicator} node `$m$'. Such a distribution leads to the formation of three regions based on the positions of the stubborn agents. 

We show that when the stubborn agents communicate with a node in Region $1$ \textit{only through $m$},  any modification $(a,b,d)$ with both $a$ and $d$ in Region $1$ becomes redundant. On the other hand, in Regions $2$ and $3$, the stubborn agents communicate with certain nodes through direct paths, making them their supporters. To identify the supporters, we further segregate the nodes into four disjoint sets: $\mathcal{T}_1$ to $\mathcal{T}_4$. The supporters of $s_1$ and $s_2$ lie in $\mathcal{T}_1$ and $\mathcal{T}_2$, respectively.
Thus, an edge modification $(a,b,d)$ where $a$ is a supporter of $s_1$ (or $s_2$) always increases its influence centrality. 
We also show that, under certain conditions, modifications with $a \in \mathcal{T}_4$ and $d \in \mathcal{T}_2$ can also increase the influence of $s_1$. Analogous results are presented for $s_2$ as well.

We extend our results to any network of type $\mathcal{C}^{1}$ by examining the special scenario where both $\mathcal{T}_2$ and $\mathcal{T}_4$ are empty. In such a case, we show that the influence centrality of $s_2$ can still be increased by an edge modification where $a$ is in $\mathcal{T}_3$ only when the degree of stubbornness of $s_2$ is $0.5$ or greater. Otherwise, such a modification can even increase the influence of $s_1$. This result highlights the significance of the position of stubborn agents in distribution according to $m$. Finally, we show that the proposed edge modifications can form a sequence that can eventually lead to the least influential stubborn agent becoming the most influential one. In future, we plan to extend our results to a wider class of strongly connected networks with multiple stubborn agents.
\vspace{-0.25cm}
\section*{Appendix A}
\noindent \textbf{\small \textcolor{black}{Proof of Theorem \ref{thm:3}
}}
We begin by establishing the relation between the underlying network and the influence centrality. 
Since the opinions evolve according to eqn. \eqref{eqn:op_model}, the final opinions satisfy $\mathbf{x}^{*}=(I_n-\beta)W\mathbf{x}^{*}+\beta\mathbf{x}(0)$. Additionally, it follows from eqn. \eqref{eqn:steady_state_op} and \eqref{eqn:influence_centrality} that average final opinion denoted by $\bar{x}$ satisfies $\bar{x}=\mathbb{1}_{n}^{T}\mathbf{x}^{*}/n=\mathbf{x}(0)^T\mathbf{c}$. We use the following equations to construct a signal flow graph $G_s=(V_s,E_s)$ \cite{mason1956feedback}. 
\begingroup
\setlength{\abovedisplayskip}{2pt}
\setlength{\belowdisplayskip}{2pt}
\begin{align}
\label{sfg:eqn}
        \mathbf{x}^{*}&=
        (I_n-\beta)W\mathbf{x}^{*}+ \beta \mathbf{x}(0) \nonumber \\
      \bar{x} &=  \mathbb{1}_{n}^{T} \mathbf{x}^{*}/n
\end{align}
\endgroup

\textbf{SFG Construction:}  $G_s$ has $n+3$ nodes where the nodes $\{1,...,n\}$ are each associated with final opinion $x^{*}_i$ of the agent $i$, the nodes $n+1$ and $n+2$ are associated with initial opinions of stubborn agents $s_1$ and $s_2$, respectively, and the node $n+3$ is associated with $\bar{x}$. The branches and their weights are derived using the eqn. \eqref{sfg:eqn}. A branch $(j,i)$ exists, if the state associated with a node $i$ depends on the state associated with a node $j$. For example: A branch $(j,i) \in E_s$ has branch gain $g_{i,j}=(1-\beta_i)w_{ij}$ for $i,j \in \{1,2,...,n\}$. Similarly, the remaining branches can be derived (refer to \cite{aashiIC} for a detailed description). It follows from eqn. \eqref{sfg:eqn} that nodes $n+1$ and $n+2$ form sources (no incoming edges) and are henceforth {denoted as $S_1$ and $S_2$}. Node $(n+3)$ is the sink (no outgoing edges) and is denoted as $O$. It follows from \eqref{sfg:eqn}, the in-degree of each node (except sources) is equal to $1$. 
In \cite{aashiIC}, the authors show that the gain of the SFG for source $S_i$ (taken one at a time) on sink $O$ is equal to the influence centrality of the corresponding stubborn agent. 

\begin{figure}[h]
    \centering
\includegraphics[width=1\linewidth,height=1.8cm,keepaspectratio]{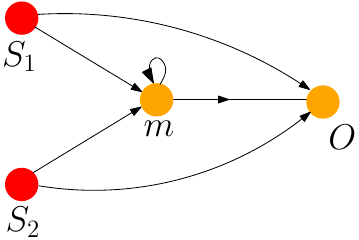}
    \caption{Reduced graph $G_s^1$}
    \label{fig:reduced_sfgs}
\end{figure}

\textbf{Reduction of $\mathbf{G_s}$:} Next, $G_s$ is reduced to a directed acyclic graph (DAG) by the index-residue reduction \cite{aashiIC,mason1956feedback} to simplify the evaluation of its gain because the gain of a DAG is simply the sum of its forward path gains.
If any node $k$ in $G_s$, has a self-loop, then it is replaced and each outgoing edge from $k$ is multiplied with $1/(1-g_{k,k})$ to obtain graph $\bar{G}_s$ \cite{mason1956feedback}. The modified graph $\bar{G}_s$ gets further reduced to $G_s^1$ as shown in Fig. \ref{fig:reduced_sfgs} by the index-residue reduction when the underlying network $\mathcal{G}$ is of type $\mathcal{C}^{1}$ \cite{aashiIC}.
\vspace{-0.3cm}
The reduced graph ${G}_s^1$ has $4$ nodes remaining from $G_s$: the sources $S_1,S_2$, the sink $O$ and the global communicator $m$ in $\mathcal{G}$. A branch $(i,j)$ exists in ${G}_s^1$ if a path from $i$ to $j$ exits in $\bar{G}_s$ (or $\mathcal{G}$) that does not pass through any other remaining nodes and 
its branch gain $g_{j,i}^{1}$ is equal to the sum of path gains of all such paths in $\bar{G}_s$. For example: the branch gain $g_{O,S_1}$ is the sum of paths in $\bar{G}_s$ from $S_1$ to $O$ that do not pass through $S_2$ or $m$.
The gain of $G_s^1$ for a source $S_i$ at sink $O$ is given by:
\begingroup
\setlength{\abovedisplayskip}{2pt}
\setlength{\belowdisplayskip}{2pt}
\begin{align}
\label{eqn:gain_eqn}
   c_{s_i}=\frac{g_{m,S_i}^1g_{O,m}^1}{1-g_{m,m}^1}+g_{O,S_i}^1
\end{align}
\endgroup
where $c_{s_i}$ is the influence centrality of $s_i$ for $i\in\{1,2\}$.

\textbf{Impact of modifications on $G_s^1$}: Having developed a relation between the influence centrality and the underlying network, we prove that an edge modification $(a,b,d)$ under the conditions in Theorem \ref{thm:3} is redundant. To achieve this, we show that the branch gain of the branches in ${G}_s^1$ remains unchanged despite edge modification $(a,b,d)$. 

The branch gain $g_{i,j}^1$ of $(i,j)$ in $G_s^1$ is the sum of direct paths from $i$ to $j$ in $\bar{G}_s$ because the other nodes remaining in $G_s^1$ form sources and sink.
In condition (1), both $a$ and $d$ are in $\mathcal{T}_4$, so neither of the sources ($S_1$ and $S_2$) have a direct path to them. Consequently, the branch gain of branches $(S_i,j)$ in $G_s^1$ for $i\in\{1,2\}$ and $j \in \{O,m\}$ do not get changed by  
the modification $(a,b,d)$. 
It follows from eqn. \eqref{eqn:gain_eqn} that the influence centrality
depends only on the branches originating from $m$. Additionally, we know that the influence centrality satisfies
$c_{s_1}+c_{s_2}=1$, therefore, $\sum_{i=1,2}({g_{m,S_i}^1g_{O,m}^1})/({1-g_{m,m}^1})+g_{O,S_i}^1=1$ holds both before and after modification. Let $\tilde{g}_{i,j}^1$ denote the branch gain of $(i,j)$ in the modified $G_s^1$. 
Thus, $(g_{m,S_1}^1+g_{m,S_2}^1)\big({g_{O,m}^1}/({1-g_{m,m}^1})-{\tilde{g}_{O,m}^1}/({1-\tilde{g}_{m,m}^1})\big)=0$. Since $(g_{m,S_1}^1+g_{m,S_1}^2) \neq 0$, it implies ${g_{O,m}^1}/({1-g_{m,m}^1})={\tilde{g}_{O,m}^1}/({1-\tilde{g}_{m,m}^1})$. Therefore,  the influence centrality remains unchanged.

In condition (2), node $a$ has a direct path from $S_1$, so the branch gain of out-going branches of $S_1$ in $G_s^1$ can get altered by $(a,b,d)$. 
The following lemma is useful in determining the impact of such modifications. 
\begin{lemma}
\label{lm:sum_direct_paths}
Consider a network $\mathcal{G}$ of type $\mathcal{C}^{1}$ graphs with two stubborn agents. In the SFG $\bar{G}_s$ derived from $\mathcal{G}$, the summation of path gains of direct paths 
\begin{itemize}
    \item from $S_i$ to a node $e$ is $\leq \frac{\beta_{s_i} (1-g_{e,e})}{1-g_{s_i,s_i}}$ where $s_i$ is a node in $\bar{G}_s$ associated with final opinion of stubborn agent $s_i$ and $i \in\{1,2\}$. Equality holds only 
    when each simple path from $m$ to $e$ passes through $s_i$.
\item from $e$ to $f$ is $\leq \frac{1-g_{f,f}}{1-g_{e,e}}$ where equality holds only if each simple path from $m$ 
to $f$ passes through $e$.
\end{itemize}
\end{lemma}
The proof of Lemma \ref{lm:sum_direct_paths} is in Appendix B.  
Using Lemma \ref{lm:sum_direct_paths}, we derive
the change in branch gain of a branch $(r,s)$ in $G_s^{1}$,  which is denoted by $\Delta g_{s,r}^1=\tilde{g}_{s,r}^1-g_{s,r}^1$.  
In terms of paths from  $S_i$ to $j$ in $\bar{G}_s$, the change $\Delta g_{j,S_i}^1$  is given as $\Delta g_{j,S_i}^1=\frac{\beta_{s_i}}{(1-g_{s_i,s_i})}\big(\sum_{P_{j,S_i}^{(a,b)}} \frac{g_{n_1,s_i}g_{n_2,n_1}...\tilde{g}_{b,a}...g_{j,n_r}}{(1-g_{n_1,n_1})...(1-g_{n_r,n_r})}+ \sum_{P_{j,S_i}^{(d,b)}} \frac{g_{n_1,s_i}g_{n_2,n_1}... (\tilde{g}_{b,d}-{g}_{b,d})...g_{j,n_k}}{(1-g_{n_1,n_1})...(1-g_{n_k,n_k})}\big)$, where $i\in\{1,2\}$, $j \in\{m,$ $O\}$, $P_{u,v}^{(x,z)}$ denotes the set of direct paths in $\bar{G}_s$ from $v $ to $ u$ passing through $(x,z)$ and $u,v,x$ and $z$ are nodes in $\bar{G}_s$, $\tilde{g}_{b,a}$ and $\tilde{g}_{b,d}$ are branch gains of $(a,b)$ and $(d,b)$ in the modified $\bar{G}_s$ and $n_1,n_2,n_k,n_r$ are nodes in $\bar{G_s}$. By definition of $(a,b,d)$, we know that $\tilde{g}_{b,a}=w$ and $(g_{b,d}-\tilde{g}_{b,d})=w$.
Since each path in sets $P_{j,S_i}^{(a,b)}$ and $P_{j,S_i}^{(d,b)}$ passes through $b$, thus the path beyond $b$ is common in both the sets and $\Delta g_{j,S_i}^1$ depends upon the direct paths from $S_1$ to $b$ only \cite{aashiIC}. Thus, $\Delta g_{j,S_i}^1$ simplifies to $\Delta g_{j,S_i}^1= K\frac{\beta_{s_i}}{(1-g_{s_i,s_i})}\big(\sum_{P_{j,S_i}^{(a,b)}} \frac{g_{n_1,s_1}g_{n_2,n_1}...g_{a,n_l}}{(1-g_{n_1,n_1})...(1-g_{a,a})} -\sum_{P_{j,S_i}^{(d,b)}} \frac{g_{n_1,s_1}g_{n_2,n_1}... g_{d,n_h}}{(1-g_{n_1,n_1})...(1-g_{d,d})}   \big)w=Kw(\frac{\mathcal{S}_{a,S_i}}{(1-g_{a,a})}-\frac{\mathcal{S}_{d,S_i}}{(1-g_{d,d})})$ where $K$ is the sum of direct paths from $b $ to $ j$ and $\mathcal{S}_{t,S_i}$ denotes the sum of direct paths from $S_i$ to $t$ in $\bar{G}_s$.

Now, we show that $(a,b,d)$ under condition (2) is redundant.
When both $a$ and $d$ are in $\mathcal{T}_1$, then the branch gains of branches in $G_s^{1}$ from $S_2$ do not get affected. Additionally, since each simple path from $m$ to $a$ and $d$ passes through $s_1$, we know from Lemma \ref{lm:sum_direct_paths} that $a$ and $d$ satisfy $\mathcal{S}_{a,S_1}=\beta_{s_1}(1-g_{a,a})/(1-g_{s_1,s_1})$ and $\mathcal{S}_{d,S_1}=\beta_{s_1}(1-g_{d,d})/(1-g_{s_1,s_1})$, respectively. It results in $\Delta g_{j,S_1}^1=0$ for $j\in \{m,O\}$. Since branches from both $S_1$ and $S_2$ are unaffected, it follows from the proof of condition (1) that influence centrality remains unchanged. The proofs for the remaining cases are also on similar lines and are omitted for brevity. $\qed$

\noindent \textbf{\small \textcolor{black}{Proof of Theorem \ref{thm:4}
}}:
Like Theorem \ref{thm:3}, we determine the impact of $(a,b,d)$ on the reduced graph $G_s^1$ for the given conditions. Under C1, $a\in \mathcal{T}_1$ and each direct path from $S_1$ to $a$ passes through $s_1$ resulting in $\mathcal{S}_{a,S_1}=\beta_{s_1} ({1-g_{a,a}})/({1-g_{s_1,s_1}})$. On the other hand, when node $d$ is in $\mathcal{T}_2$ or $\mathcal{T}_4$, then $S_1$ does not have a direct path to $d$. So, $\mathcal{S}_{d,S_1}=0$. When $d \in \{\mathcal{T}_1,\mathcal{T}_3\}$, we know from Lemma \ref{lm:sum_direct_paths} that $\mathcal{S}_{d,S_1}<\beta_{s_1} ({1-g_{d,d}})/({1-g_{s_1,s_1}})$ under conditions in C-1. Therefore, $\Delta g_{j,S_1}^1>0$ for $j \in \{m,O\}$. In $G_s^{1}$, the branch gain of branches from $S_1$ increases and from $S_2$ remains unchanged (except when $d\in\mathcal{T}_2$ where it decreases). 
If after modification, branch gain of branches from $m$ varies such that $g_{O,m}^1/(1-g_{m,m}^1)$ increases, then it follows from eqn. \eqref{eqn:gain_eqn} that $c_{s_1}$ increases after modification.  
On the other hand, if $g_{O,m}^1/(1-g_{m,m}^1)$ decreases, then $c_{s_2}$ reduces and consequently $c_{s_1}$ increases. Therefore, $c_{s_1}$ increases independent of change in branch gains of $(m,m)$ and $(m,O)$. 

Under the condition C2, since $a \in \mathcal{T}_4$, we have $\mathcal{S}_{a,S_i}=0$ for both $S_1$ and $S_2$. On the other hand, since $d \in \mathcal{T}_2$, thus $\mathcal{S}_{d,S_2}$ has a positive value. As a result, $\Delta g_{j,S_2}^{1}<0$ for $j \in \{m,O\}$. Since $S_1$ does not have a direct path to $d$, the branches $(S_1,j)$ remain unaffected. Therefore, it follows from the above analysis that $c_{s_1}$ increases. A similar analysis shows that conditions in C1 and C2 pertaining to $s_2$ increase its influence centrality.
$\qed$

\noindent \textbf{\small \textcolor{black}{Proof of Corollary \ref{Cor:stubbornness}
}}: Under the given conditions, the branch gains of all branches in $G_s^1$ vary due to $(a,b,d)$. First, consider the branches $(S_2,j)$ for $j\in\{m,O\}$. It follows from Lemma \ref{lm:sum_direct_paths} that $\Delta g_{j,S_2}^{1}>0$. Now, we consider the branches from $S_1$ (\textit{i.e} $(S_1,j)$). When $d \in \mathcal{T}_4$, then $\mathcal{S}_{d,S_1}=0$ resulting in $\Delta g_{j,S_1}^{1}>0$. 
Similarly, when $d \in \mathcal{T}_1$ or $\mathcal{T}_3$, 
the branch gain $(S_1,j)$ increases when $\mathcal{S}_{a,S_1}>\mathcal{S}_{d,S_1}$. 
Under the present scenario, the gains of the branches for both sources increase but the influence centrality of only one stubborn agent can increase. Thus, it follows from eqn. \eqref{eqn:gain_eqn} that 
$c_{s_i}$ increases if the increase in the branch gains from $S_i$ is higher. 

Now, we determine the conditions under which $c_{s_2}$ always increases. 
We begin by maximising $\Delta g_{j,S_1}^1$ and then determine the conditions under which it is still less than $\Delta g_{j,S_2}^1$. To maximise $\Delta g_{j,S_1}^1$, we increase $\mathcal{S}_{a,S_1}$ to as high and reduce $\mathcal{S}_{d,S_1}$ to as low as possible.
First, we maximise $\mathcal{S}_{a,S_1}$.
Under given condition, each simple path from $m$ (equivalently $S_1$) to $a$ passes through $s_2$, as a result $\mathcal{S}_{a,S_1}=\mathcal{S}_{s_2,S_1}\mathcal{S}_{a,s_2}$.
From Lemma \ref{lm:sum_direct_paths}, it follows that $\mathcal{S}_{s_2,S_1}$ is maximum if every path from $m$ to $s_2$ passes through $s_1$. It is equal to $\beta_{s_1} \frac{1-g_{s_{2},s_2}-\beta_{s_2}}{1-g_{s_1,s_1}}$ where $-\beta_{s_2}$ is introduced in the numerator because branch $(S_2,s_2)$ in $\bar{G}_s$ with edge weight $\beta_{s_2}$ is not present in paths from $S_1$ to $s_2$. Further, $S_1$ is least connected to $d$ if it either does not have a direct path to $d\in \mathcal{T}_4$ (here, $\mathcal{S}_{d,S_1}=0$) or when each path from $S_1$ to $d$ also passes through $s_2$ (here also, $\mathcal{S}_{d,S_1}=\mathcal{S}_{s_2,S_1}\mathcal{S}_{d,s_2}$). We consider the latter case, and it can be shown that the former also equivalently follows. In the latter case, $\Delta g_{j,S_1}^{1}=Kw\mathcal{S}_{s_2,S_1}(\frac{\mathcal{S}_{a,s_2}}{1-g_{a,a}}-\frac{\mathcal{S}_{d,s_2}}{1-g_{d,d}})$. Comparing it with  $\Delta g_{j,S_2}^{1}$, we get: 
$\Delta g_{j,S_1}^1-\Delta g_{j,S_2}^1=Kw \big(\mathcal{S}_{s_2,S_1}(\frac{\mathcal{S}_{a,s_2}}{1-g_{a,a}}-\frac{\mathcal{S}_{d,s_2}}{1-g_{d,d}})-(\frac{\mathcal{S}_{a,S_2}}{1-g_{a,a}}-\frac{\mathcal{S}_{d,S_2}}{1-g_{d,d}})\big)=Kw \big(\frac{\mathcal{S}_{s_2,S_1}}{\beta_{s_2}}-1\big)(\frac{\mathcal{S}_{a,S_2}}{1-g_{a,a}}-\frac{\mathcal{S}_{d,S_2}}{1-g_{d,d}})$.
Here, the second inequality follows because $\mathcal{S}_{k,S_i}=\beta_{s_i}\mathcal{S}_{k,s_i}$ in $\bar{G}_s$. Now, if $\mathcal{S}_{s_2,S_1}/{\beta_{s_2}}-1<0$, then $c_{s_2}$ increases and $c_{s_1}$ reduces. It implies that if $\frac{\beta_{s_1}(1-g_{s_2,s_2}-\beta_{s_2})}{\beta_{s_2}(1-g_{s_1,s_2})}<1$ holds, then $c_{s_2}$ increases. Now, when both $s_1$ and $s_2$ do not have self-loops, this condition simplifies to $\beta_{s_1}<\beta_{s_2}/(1-\beta_{s_2})$ which always hold for $\beta_{s_2}>0.5$.$\qed$
\vspace{-0.25cm}
\section*{Appendix B}
\noindent \textbf{\small \textcolor{black}{Proof of Lemma \ref{lm:sum_direct_paths}
}}: 
Let the nodes in $\mathcal{G}$ be distributed according to $m$ as defined in Lemma \ref{lemma:distribution_of_nodes}. Consider 
a node $e\in \mathcal{T}_1$ where each direct path from $m$ to $e$ passes through $s_1$. 
Since $e$ is of type $\mathcal{T}_1$, each node on a direct path from $S_1$ to $e$ is of type $\mathcal{T}_1$. Under the given conditions,
a node $j$ (different than $s_1$) on a direct path from $S_1$ to $e$ satisfies:
\begin{itemize}
    \item if $j\in \mathcal{L}_{u+1}^{m}$, it has only $s_1$ as its in-neighbour (because each direct path from $m$ to $j$ also passes through $s_1$) resulting in $g_{j,s_1}=1-g_{j,j}$. Since $S_1$ has an outgoing edge only to $s_1$ with branch gain $\beta_{s_1}$, the path gain of direct path $S_1$ to $ s_1 $ to $ j$ equals  $\mathcal{S}_{j,S_1}=\beta_{s_1} (1-g_{j,j})/(1-g_{s_1,s_1})$.
    \item If $j\in \mathcal{L}_{u+2}^{m}$, then it has $s_1$ or node(s) from $\mathcal{L}_{u+1}^{m}$ as its neighbour(s). In this scenario,
$\mathcal{S}_{j,S_1}=\frac{\beta_{s_1}}{({1-g_{s_1,s_1}})}$ $\big(\sum_{k\in N_{j}^{in}\setminus\{s_1,j\}}\frac{g_{j,k}g_{k,s_1}}{1-g_{k,k}}+g_{j,s_1}\big)=\frac{1-g_{j,j}}{1-g_{s_1,s_1}}$,
where the second equality holds because each in-neighbour $k$ satisfies ${g_{k,s_1}}/({1-g_{k,k}})=1$.
\item By induction, if $j \in \mathcal{L}_{u+h}^{m}$ such that it lies on a direct path $m $ to $ e$ 
Then, node $j$ also satisfies $\mathcal{S}_{j,S_1}=({1-g_{j,j}})/({1-g_{s_1,s_1}})$.
\end{itemize}
However, when node $e$ has a direct path from $m$ that does not pass through $s_1$, a node (say $r$) exists
on a direct path $S_1 $ to $ e$ 
which has a node (say $t\in \mathcal{T}_4$) as its in-neighbour. 
Let each in-neighbours of $r$ except $t$ have all direct paths from $m$ passing through $s_1$, then, 
$\mathcal{S}_{r,S_1}=\beta_{s_1}(\sum_{k\in N_{r}^{in}\setminus\{t\}}g_{r,k}\mathcal{S}_{k,S_1}/(1-g_{k,k}))=\beta_{s_1}(\sum_{k\in N_{r}^{in}\setminus\{t\}}g_{r,k})/(1-g_{s_1,s_1})<\beta_{s_1} (1-g_{r,r})/(1-g_{s_1,s_1})$, where the last inequality holds because $t$ does not lie on paths from $S_1$ to $e$. It is simple to show that this holds for every node on path $r $ to $ e$.
A similar analysis proves the remaining conditions.$\square$


\vspace{-0.2cm}
\bibliographystyle{ieeetr} 
\bibliography{references}





\end{document}